\documentclass[aps,prl,twocolumn,unsortedaddress,superscriptaddress,
showpacs]{revtex4-1}

\usepackage{amsmath,amssymb,amsbsy}
\usepackage{enumerate,graphicx,multirow,dcolumn}
\usepackage[capitalize]{cleveref}
\usepackage{xstring}
\usepackage{relsize}

\usepackage[usenames]{color}
\usepackage[normalem]{ulem}

\newcommand{\ket}[1]{\left|#1\rangle\right.}

\newcommand{\matrixel}[3]{\langle#1\vphantom{#2#3}|#2|#3\vphantom{#1#2}
\rangle}
\newcommand{\ev}[1]{\langle#1\rangle}

\renewcommand{\vec}[1]{\mathbf{#1}}
\newcommand{\gvec}[1]{\boldsymbol{#1}}

\begin{document}
\title{Quantum Monte Carlo Calculations of Light Nuclei Using Chiral 
Potentials}

\author{J.~E.~Lynn}
\email[E-mail:~]{joel.lynn@gmail.com}
\affiliation{Theoretical Division, Los Alamos National Laboratory,
Los Alamos, New Mexico 87545, USA}
\author{J.~Carlson}
\affiliation{Theoretical Division, Los Alamos National Laboratory,
Los Alamos, New Mexico 87545, USA}
\author{E.~Epelbaum}
\affiliation{Institut f\"ur Theoretische Physik II, Ruhr-Universit\"at
Bochum, 44780 Bochum, Germany}
\author{S.~Gandolfi}
\affiliation{Theoretical Division, Los Alamos National Laboratory,
Los Alamos, New Mexico 87545, USA}
\author{A.~Gezerlis}
\affiliation{Department of Physics, University of Guelph,
Guelph, Ontario, N1G 2W1, Canada}
\author{A.~Schwenk}
\affiliation{Institut f\"ur Kernphysik,
Technische Universit\"at Darmstadt, 64289 Darmstadt, Germany}
\affiliation{ExtreMe Matter Institute EMMI, GSI Helmholtzzentrum f\"ur
Schwerionenforschung GmbH, 64291 Darmstadt, Germany}

\begin{abstract}
We present the first Green's function Monte Carlo calculations of light
nuclei with nuclear interactions derived from chiral effective field 
theory up to next-to-next-to-leading order.
Up to this order, the interactions can be constructed in a local form
and are therefore amenable to quantum Monte Carlo calculations.
We demonstrate a systematic improvement with each order for the binding
energies of $A=3$ and $A=4$ systems.
We also carry out the first few-body tests to study perturbative
expansions of chiral potentials at different orders, finding
that higher-order corrections are more perturbative for softer 
interactions.  Our results confirm the necessity of a 
three-body force for correct reproduction of experimental 
binding energies and radii, and pave the
way for studying few- and many-nucleon systems using quantum Monte Carlo
methods with chiral interactions.
\end{abstract}
\maketitle

Important advances in our knowledge of light nuclei have
been possible in recent years by using sophisticated numerical 
techniques like hyperspherical harmonics, the no-core shell model, and
the Green's function Monte Carlo (GFMC) method.
In particular, the nuclear GFMC method is one of the most accurate
methods used to calculate the ground and excited state energies and
other properties of light nuclei with mass number $A\le12$ by using
realistic nuclear Hamiltonians~\cite{carlson1987,pudliner1997,
wiringa2000,pieper2001.1,pieper2002,lovato2013,lovato2014} based on the
Argonne $v_{18}$ two-body potential~\cite{wiringa1995} and the
Urbana/Illinois models of three-body forces~\cite{pudliner1995,
pieper2001.2}.
Despite the many successes of the nuclear GFMC method, until now it has
been limited to modern phenomenological potentials.
Interactions derived from chiral effective field theory 
(EFT)~\cite{epelbaum2009,machleidt2011} provide a direct connection 
between \textit{ab initio} nuclear structure calculations and the 
underlying theory of strong interactions, quantum chromodynamics (QCD).
These potentials have been successfully used in various regions
of the nuclear landscape: from structure and reactions of light nuclei
~\cite{epelbaum2010,barrett2013,bacca2013} to medium-mass 
nuclei~\cite{otsuka2010,hergert2012,binder2013,hagen2014.2,soma2013,
wienholtz2013}
to infinite matter~\cite{hebeler2010.1,hebeler2010.2,gezerlis2013,
kruger2013,hagen2014}.
In this work, we combine, for the first time, the accurate nuclear 
GFMC machinery with chiral EFT interactions,
which makes possible the first few-body studies of higher-order
corrections in the chiral expansion.

The GFMC method is an exact method for studying nuclei with chiral
interactions, because it works with the interactions in their bare form;
that is, the Hamiltonian does not need to be softened by using 
renormalization group or other techniques~\cite{bogner2010}.
Therefore, GFMC calculations of light nuclei with chiral EFT 
interactions will also be important to benchmark calculations using
other methods that rely on such techniques.
Until recently, nucleon-nucleon (NN) interactions 
derived from chiral EFT have been nonlocal, a feature which naturally
results from the construction of these interactions in momentum space
where locality is not typically an important consideration.
For many nuclear structure methods, nonlocal interactions do not pose 
any problems.
In the case of the GFMC method, however, nonlocality poses nontrivial 
technical challenges.
Sources of nonlocality in chiral EFT include the regulator choices, 
momentum-dependent contact interactions, and higher-order pion exchanges
and relativistic contributions.
The latter two appear only at next-to-next-to-next-to-leading order 
(N$^3$LO) and beyond.
Up to next-to-next-to-leading order (N$^2$LO), the other two sources can
be eliminated by choosing local regulators and an appropriate set of
contact interactions as discussed in Ref.~\cite{gezerlis2013}.
This opens up GFMC calculations of light nuclei with chiral potentials.

In this Letter, we discuss the first GFMC calculations of light nuclei 
for $A\le4$ using NN interactions derived from chiral EFT.
We present a systematic study of the ground-state energies at leading 
order (LO), next-to-leading order (NLO), and N$^2$LO and study the
cutoff dependence at each order.
We first briefly review the GFMC method and discuss the interaction
used herein.
Then we present our results for the $A\le4$ systems and discuss the
perturbative expansion of these forces at different values of the
regulator cutoff.

The GFMC method consists of propagating in imaginary time $t$ a 
trial wave function $\ket{\Psi_T}$ to extract the ground-state wave 
function $\ket{\Psi_0}$.
In the long imaginary-time limit, one has
\begin{equation}
\lim_{t\rightarrow\infty}e^{-Ht}\ket{\Psi_T}\rightarrow\ket{\Psi_0},
\end{equation}
with $H$ the Hamiltonian of the system, if $\ket{\Psi_T}$ is not
orthogonal to $\ket{\Psi_0}$.
Ground-state and low-lying excited-state observables are calculated by 
stochastic integration of the matrix elements 
$\matrixel{\Psi_T}{Oe^{-Ht}}{\Psi_T}$,
with $O$ some observable.
For reviews of the method, see, for example, 
Refs.~\cite{pudliner1997,pieper2001.1}.
For the sampling of the propagator, $e^{-Ht}$, the standard GFMC 
method relies on locality of the potential.
Though some progress has been made on this
front~\cite{lynn2012,lynn2013}, it remains technically challenging to 
sample nonlocal terms by using the GFMC method without introducing large
statistical errors.
Local chiral EFT interactions allow for the use of the GFMC method
with a minimum of further complications to calculate the propagator;
however, a careful optimization of the two-body correlations which
enter the wave function is necessary to account for the new potentials
(as these are considerably different from the harder Argonne
family of potentials).
An attempt to develop a quantum Monte Carlo method to deal with 
nonlocal nuclear forces has been presented in Ref.~\cite{roggero2014}
using the soft N$^2$LO potential of Ref.~\cite{ekstrom2013}.
Auxiliary-field quantum Monte Carlo calculations for a chiral 
interaction with a sharp cutoff were recently presented in 
Ref.~\cite{wlazlowski2014}.

We first clarify the notions of local and nonlocal interactions.
If $\vec{p}=(\vec{p}_1-\vec{p}_2)/2$ and 
$\vec{p}^\prime=(\vec{p}_1^\prime-\vec{p}_2^\prime)/2$ are the
incoming and outgoing relative momenta of the nucleon pair,
respectively, it is convenient to work in terms of the momentum transfer 
$\vec{q}=\vec{p}^\prime-\vec{p}$ and the momentum transfer in the 
exchange channel $\vec{k}=(\vec{p}^\prime+\vec{p})/2$.
When Fourier transformed, terms with $\vec{q}$ lead 
to local interactions that depend only on the interparticle distance 
$\vec{r}$.
However, terms with $\vec{k}$ are nonlocal 
contributions depending on $\gvec{\nabla}_\vec{r}$ that complicate 
the sampling of the propagator in the GFMC method.
The only exception to this is the spin-orbit term, which contains
a $\vec{q}\times\vec{k}$ term that can be included in the GFMC
propagator~\cite{pudliner1997}.

Chiral EFT provides a systematic expansion for nuclear forces and
predicts a hierarchy of two- and many-nucleon
interactions~\cite{epelbaum2009,machleidt2011}. 
At a given order, the interactions receive contributions from pion
exchanges, which make up the long- and intermediate-range parts, as
well as from short-range contact interactions. 
In particular, up to N$^2$LO, the unregulated one- and two-pion-exchange 
contributions~\cite{kaiser1997,epelbaum2004} are local. 
To construct local chiral potentials, 
Refs.~\cite{gezerlis2013,gezerlis2014} regulated the pion-exchange 
contributions with a regulator directly in coordinate space 
$f_\text{long}(r)=1-e^{-(r/R_0)^4}$, where $R_0$ is a cutoff. 
We use $R_0$ of 1.0, 1.1, and 1.2~fm, which approximately 
correspond to momentum cutoffs 500, 450, and 400~MeV, 
respectively. These values are obtained by Fourier transforming the
regulator function, integrating over all momenta, and identifying the
result with a sharp cutoff~\cite{gezerlis2014}.
In addition, following Ref.~\cite{epelbaum2004}, we employ for the
two-pion-exchange contributions, a spectral-function regularization with
cutoff $\tilde{\Lambda}$ (we will use $\tilde{\Lambda}=1000$~MeV). 
The dependence on $\tilde{\Lambda}$ is very weak~\cite{gezerlis2014}, 
as we will demonstrate comparing to results for 
$\tilde{\Lambda}=1400$~MeV.
For the short-range interactions, the local chiral potentials of
Refs.~\cite{gezerlis2013,gezerlis2014} select from the overcomplete
set of operators ones that are local in coordinate space. 
This is possible up to N$^2$LO; at N$^3$LO, a number of nonlocal 
interactions will survive. 
For these higher-order interactions, we expect that they
can be included perturbatively in the GFMC calculation, as is done
with nonlocal parts in the Argonne $v_{18}$ potential 
already~\cite{pudliner1997}. 
The short-range interactions are then regulated with a regulator
$\sim e^{-(r/R_0)^4}$ complementary to the long-range one.

\begin{figure}[t]
\includegraphics{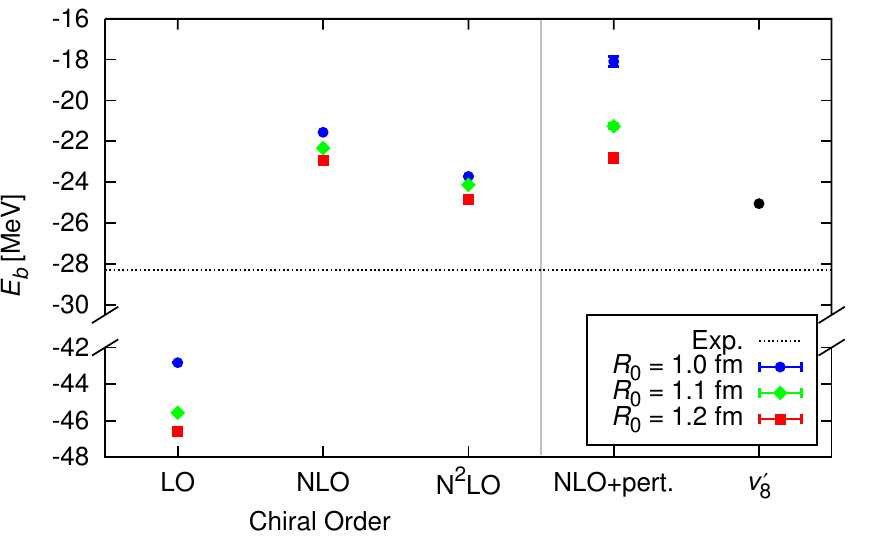}
\caption{(color online).\label{fig:he4}
$^4$He binding energies ($E_b$) at LO, NLO, and N$^2$LO compared with 
experiment and with the Argonne $v_8^\prime$ energy.
Also shown is a first-order perturbation-theory calculation of the 
N$^2$LO binding energy using the wave function at NLO: 
$E_\text{NLO}+V_\text{pert}$.
See~\cref{eq:vpert} and the discussion that follows.
The GFMC statistical errors are generally smaller than the points.}
\end{figure}

Interactions derived from chiral EFT are expected to show an 
order-by-order improvement or convergence.
However, note that a calculation at N$^2$LO with
only two-body forces is incomplete, as three-body forces enter at this
order.
For each nucleus, we perform calculations at LO, NLO, and N$^2$LO,
varying the cutoff $R_0$ from 1.0 to 1.2~fm.
In \cref{fig:he4} and \cref{tab:he4,tab:he3,tab:h3}, we present the GFMC
results for the binding energies of the $A=3$ and $A=4$ nuclei for the
various chiral potentials (the Coulomb potential is also included).
As the chiral order increases, we can see a reduction in the 
theoretical uncertainty coming from the $R_0$ variation.
For example, for $^4$He, the bands are $\sim3.8$~MeV, $\sim1.4$~MeV, 
and $\sim1.1$~MeV at LO, NLO, and N$^2$LO, respectively.
For $^4$He at N$^2$LO, we also used the spectral-function cutoff of 
$\tilde{\Lambda}=1400$~MeV with $R_0=1.0$~fm and $R_0=1.2$~fm.
These calculations lowered the $^4$He binding energy by 0.41~MeV
$(\sim2\%)$ and 0.46~MeV $(\sim2\%)$, respectively (compared to
the case with $\tilde{\Lambda}=1000$~MeV), which demonstrates a weak
dependence on $\tilde{\Lambda}$.
The calculated radii are consistent with the general trend in the 
binding energies: that is, at LO, the nuclei are significantly 
overbound, and the corresponding radii are too small compared to 
experiment; at NLO, the nuclei are underbound, and the 
radii are larger than experiment; at N$^2$LO, the nuclei are still
underbound, but closer to experiment, and the corresponding radii are
smaller (closer to experiment).
\begin{table}[hb]
\caption{\label{tab:he4}Binding energies and point proton radii for 
$^4$He.
The errors given are statistical GFMC uncertainties.
The experimental binding energy and root-mean-square (rms) point proton
radius are $E_b=-28.31$~MeV and $\sqrt{\ev{r_\text{pt}^2}}=1.45$~fm,
respectively.}
\begin{ruledtabular}
\begin{tabular}{ccdc}
Order&$R_0$ [fm]&\multicolumn{1}{c}{$E_b$ [MeV]}&
$\sqrt{\ev{r^2_\text{pt}}}$ [fm]\\
\hline
\multirow{3}{*}{LO}&1.0&-42.83(1)&1.02(1)\\
&1.1&-45.57(2)&1.00(1)\\
&1.2&-46.62(1)&1.00(1)\vspace{0.25em}\\
\multirow{3}{*}{NLO}&1.0&-21.56(1)&1.57(1)\\
&1.1&-22.33(1)&1.54(1)\\
&1.2&-22.94(6)&1.53(1)\vspace{0.25em}\\
\multirow{3}{*}{N$^2$LO}&1.0&-23.72(1)&1.52(1)\\
&1.1&-24.13(1)&1.50(1)\\
&1.2&-24.86(1)&1.47(1)\\
\end{tabular}
\end{ruledtabular}
\end{table}
\begin{table}[htb]
\caption{\label{tab:he3}Binding energies and point proton radii for 
$^3$He.
The errors given are statistical GFMC uncertainties.
The experimental binding energy and rms point proton radius are
$E_b=-7.72$~MeV and $\sqrt{\ev{r_\text{pt}^2}}=1.76$~fm, respectively.}
\begin{ruledtabular}
\begin{tabular}{ccdc}
Order&$R_0$ [fm]&\multicolumn{1}{c}{$E_b$ [MeV]}&
$\sqrt{\ev{r^2_\text{pt}}}$ [fm]\\
\hline
\multirow{3}{*}{LO}&1.0&-10.42(1)&1.36(1)\\
&1.1&-10.78(1)&1.36(1)\\
&1.2&-10.88(1)&1.36(1)\vspace{0.25em}\\
\multirow{3}{*}{NLO}&1.0&-6.35(2)&1.92(2)\\
&1.1&-6.56(1)&1.90(2)\\
&1.2&-6.67(1)&1.88(1)\vspace{0.25em}\\
\multirow{3}{*}{N$^2$LO}&1.0&-6.78(1)&1.87(2)\\
&1.1&-6.90(1)&1.84(1)\\
&1.2&-7.01(1)&1.82(1)\\
\end{tabular}
\end{ruledtabular}
\end{table}
\begin{table}[bt]
\caption{\label{tab:h3}Binding energies and point proton radii for 
$^3$H.
The errors given are statistical GFMC uncertainties.
The experimental binding energy and rms point proton radius are
$E_b=-8.48$~MeV and $\sqrt{\ev{r_\text{pt}^2}}=1.59$~fm, respectively.}
\begin{ruledtabular}
\begin{tabular}{ccdc}
Order&$R_0$ [fm]&\multicolumn{1}{c}{$E_b$ [MeV]}&
$\sqrt{\ev{r^2_\text{pt}}}$ [fm]\\
\hline
\multirow{3}{*}{LO}&1.0&-11.00(1)&1.27(1)\\
&1.1&-11.42(1)&1.26(1)\\
&1.2&-11.54(1)&1.27(1)\vspace{0.25em}\\
\multirow{3}{*}{NLO}&1.0&-7.10(1)&1.62(3)\\
&1.1&-7.25(2)&1.62(3)\\
&1.2&-7.35(1)&1.64(3)\vspace{0.25em}\\
\multirow{3}{*}{N$^2$LO}&1.0&-7.55(1)&1.61(2)\\
&1.1&-7.63(1)&1.61(3)\\
&1.2&-7.74(1)&1.58(2)\\
\end{tabular}
\end{ruledtabular}
\end{table}

The LO calculations bear additional discussion since, 
as~\cref{fig:he4} and~\cref{tab:he4,tab:he3,tab:h3} 
show, we find that the nuclei are significantly overbound: by as much as 
$\sim65\%$ of the experimental binding energy in the case of $^4$He 
with the cutoff at $R_0=1.2$~fm.
In the LO case, there are only two low-energy couplings, and the phase 
shifts are fit only up to $E_\text{lab}=50$~MeV;
therefore, the effective-range physics is not reproduced and the
potential is too attractive~\cite{gezerlis2014}.
Since we expect the lightest nuclei with $A=2$ and $A=3$ to be least 
sensitive to higher energy scales, we might expect that these nuclei are 
less overbound than $^4$He.
This trend is, indeed, borne out.
At LO, $^3$He and $^3$H are overbound by as much as $\sim41\%$ and 
$\sim36\%$, respectively (compared with the $\sim65\%$ for $^4$He).
The deuteron is underbound by $\sim9\%$~\cite{gezerlis2014}.

The chiral EFT expansion is an expansion in powers of momentum or of
the pion mass $\sim Q$ over a breakdown scale $\Lambda_b$.
As we increase the chiral order, we expect suppression of the 
contributions from higher orders by powers of $Q/\Lambda_b$.
It is clear from the results presented 
in~\cref{fig:he4,tab:he4,tab:he3,tab:h3} that the NLO contribution is 
an important correction to the LO results.
But the same results suggest that the contributions from N$^2$LO are 
small relative to the NLO contributions.
There is also evidence from calculations of the neutron-matter energy
using chiral potentials that suggests perturbative behavior of chiral
interactions~\cite{gezerlis2013,kruger2013}.
Therefore, it seems reasonable to attempt first-order perturbation
theory for the $A\le4$ nuclei, treating the difference in the 
potentials as a perturbation:
\begin{equation}
\label{eq:vpert}
V_\text{pert}=V_\text{N$^2$LO}-V_\text{NLO}.
\end{equation}
The results of these calculations for $^4$He are shown 
in~\cref{fig:he4}.
For each of the three values of the cutoff, we find the first-order 
contribution to be positive.
The smallest correction comes in the $R_0=1.2$~fm case as might be
expected.
(Larger $R_0$ corresponds to lower $\Lambda$ in momentum space, so that
$R_0=1.2$~fm is the softest potential used.)
It would, of course, be desirable to compute higher-order perturbative
corrections; however, it is difficult to obtain the second-order result
or beyond.

We can, however, study first-, second-, and third-order 
perturbation-theory calculations for the deuteron.
The methods developed in Refs.~\cite{lynn2012,lynn2013} allow for the
determination of the first $N$ excited states of the deuteron.
In the calculations discussed here, $N\sim800$, giving truncation 
errors of less than $10^{-10}$~MeV.
\Cref{tab:h2pert} shows the results of these calculations with
the NLO and N$^2$LO potentials with three different cutoffs.
The first-order correction is positive and varies from
$12\%$~to~$33\%$ of the NLO deuteron binding energy for 
$R_0=1.2$--$1.0$~fm.
The corrections at second and third order are both 
negative and range from $13\% - 31\%$ (at second order) 
and from $0.46\%$~to~$0.93\%$ (at third order).
The $R_0=1.0$~fm case has the largest corrections at each order in the
perturbation expansion; the $R_0=1.2$~fm case has the smallest.
There is some evidence, then, that the perturbative expansion for
$V_\text{pert}$ is converging in each case but faster for the softer
potentials.

The perturbative check we have presented treats the difference in the
fitted potentials at N$^2$LO and NLO as a perturbation,~\cref{eq:vpert}.
We have also tested whether the new interactions entering at N$^2$LO are
perturbative.
To this end, we take the NLO parts of the N$^2$LO potential and treat 
the higher-order interactions as a perturbation.
In this approach, the deuteron and $^4$He are unbound at first order in
perturbation theory.
These results appear to be due to the large $c_i$'s which enter at 
N$^2$LO.
This pattern may be different in a chiral EFT with explicit Delta 
degrees of freedom, where the N$^2$LO $c_i$'s are natural.

\begin{table}[t]
\caption{\label{tab:h2pert}Perturbation calculations for $^2$H
using the NLO and N$^2$LO potentials.
The notation $E_\text{NLO}$ indicates the ground-state 
energy of the NLO Hamiltonian.
$V_\text{pert}^{(n)}$ indicates the sum of the perturbative corrections
up to the $n$th order.}
{\renewcommand{\arraystretch}{1.30}
\begin{ruledtabular}
\begin{tabular}{lddd}
\multirow{2}{*}{Calculation}&\multicolumn{3}{c}{$E_b$ [MeV]}\\
&\multicolumn{1}{c}{\scriptsize{$R_0\!=\!1.0$\! fm}}&
\multicolumn{1}{c}{\scriptsize{$R_0\!=\!1.1$\! fm}}&
\multicolumn{1}{c}{\scriptsize{$R_0\!=\!1.2$\! fm}}\\
\hline
$E_\text{NLO}$&-2.15&-2.16&-2.16\\
$E_\text{NLO}+V_\text{pert}^{(1)}$&-1.44&-1.80
&-1.90\\
$E_\text{NLO}+V_\text{pert}^{(2)}$&
-2.11&-2.17&-2.18\\
$E_\text{NLO}+V_\text{pert}^{(3)}$&
-2.13&-2.18&-2.19\\
$E_\text{N$^2$LO}$&-2.21&-2.21&-2.20\\
\end{tabular}
\end{ruledtabular}}
\end{table}

\begin{figure}[b]
\includegraphics{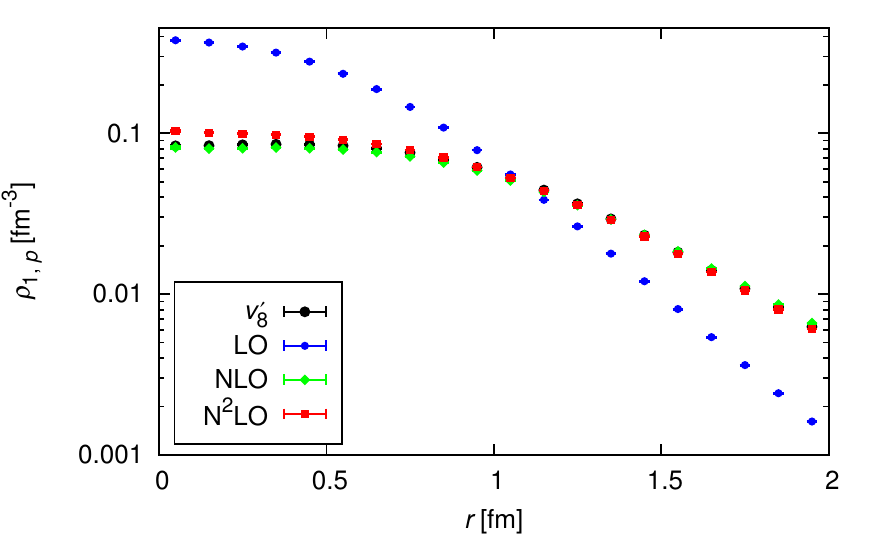}
\caption{\label{fig:1bd}(color online).
One-body proton distributions for $^4$He with $R_0=1.2$~fm at 
LO, NLO, and N$^2$LO compared with results for the Argonne $v_8^\prime$
interaction.
The error bars (generally smaller than the symbol size) are the 
statistical errors.}
\end{figure}

\begin{figure}[b!t]
\includegraphics{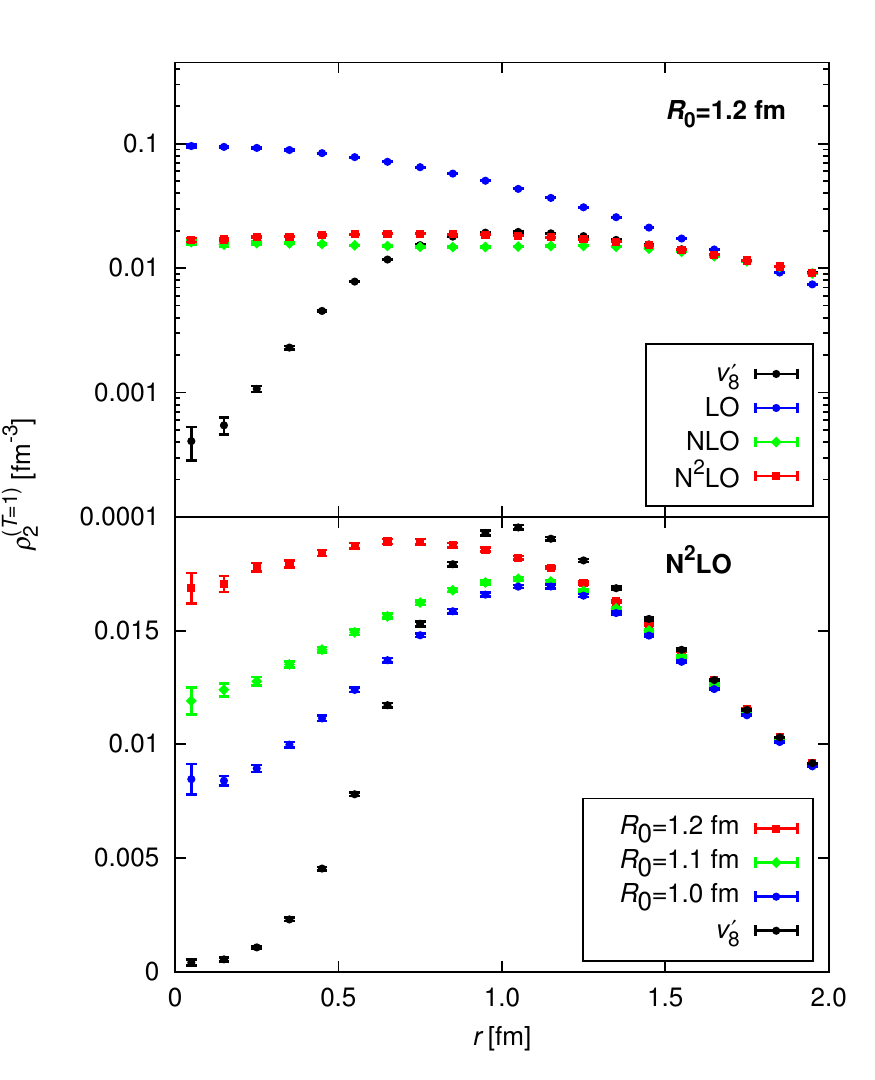}
\caption{\label{fig:2bd}(color online).
Two-body $T=1$ distributions for $^4$He using chiral potentials and the
Argonne $v_8^\prime$ interaction.
The top panel has the distributions calculated with $R_0=1.2$~fm at LO, 
NLO, and N$^2$LO.
The bottom panel shows the dependence of the N$^2$LO distribution at 
short distances on the cutoff $R_0$.}
\end{figure}

In addition to the binding energies and radii, we have calculated one-
and two-body distributions and display them in~\cref{fig:1bd,fig:2bd}.
The proton distribution is given by
\begin{equation}
\begin{split}
\rho_{1,p}(r)=\frac{1}{4\pi r^2}
\left\langle\Psi_0\vphantom{\sum}\right|\sum_i\frac{1+\tau_z(i)}{2}
\delta(r-|\vec{r}_i-\vec{R}_\text{c.m.}|)
\left|\vphantom{\sum}\Psi_T\right\rangle,
\end{split}
\end{equation}
where $\vec{r}_i$ is the position of the $i$th nucleon, 
$\vec{R}_\text{c.m.}$ is the center-of-mass of the nucleus, and 
$\tau_z(i)/2$ is the $z$ component of the isospin of the $i$th nucleon.
We have calculated the two-body distribution functions in the $T=1$ 
isospin state, defined as
\begin{equation}
\rho_2^{(T=1)}(r)=\frac{3\rho_{2,1}(r)+
\rho_{2,\gvec{\tau}\cdot\gvec{\tau}}(r)}{4},
\end{equation}
where
\begin{equation}
\rho_{2,O}(r)=\frac{1}{4\pi r^2}
\left\langle\vphantom{\sum}\Psi_0\right|
\sum_{i<j}O_{ij}\delta(r-|\vec{r}_{ij}|)
\left|\vphantom{\sum}\Psi_T\right\rangle.
\end{equation}
In~\cref{fig:1bd,fig:2bd}, it is clear that for distances 
$r\gtrsim1.5$~fm the NLO, N$^2$LO, and Argonne $v_8^\prime$
distributions are very similar.
At short distances ($r\lesssim1.5$~fm) the LO distributions are 
significantly larger than the distributions calculated with the other 
interactions.
In~\cref{fig:2bd}, the different short-range behavior of the chiral 
forces and the Argonne $v_8^\prime$ interaction is clear;
the softer two-body $T=1$ NLO and N$^2$LO distributions (larger values 
of the distributions at the origin) suggest that short-range 
correlations between nucleons reflect the presence or absence of a
hard core in the interaction~\cite{gezerlis2014}.
In the lower panel in~\cref{fig:2bd}, we show the dependence of 
$\rho_2^{(T=1)}(r)$ on the cutoff by using the N$^2$LO potentials.
The one- and two-body distributions lend further support to the
discussion above about the overbinding of the nuclei at LO.
\Cref{fig:1bd,fig:2bd} imply that at LO the nucleons tend to be closer 
together than at higher order or with the phenomenological 
Argonne~$v_8^\prime$ potential.

We have presented a systematic GFMC study of light nuclei $A\le4$ 
with local NN interactions derived from chiral EFT up to N$^2$LO.
There is an order-by-order improvement for the binding energies, which
is also shown by the weaker cutoff dependence.
Our calculations confirm the necessity of a three-body force for nuclei
with $A\ge3$.
We have also presented the first nonperturbative study of the interplay
between different orders in the chiral expansion.
We find that higher-order contributions are more perturbative for the 
softer potentials, and our calculations for the deuteron 
suggest that the perturbation expansion is converging to the result at 
N$^2$LO.
This study lays the groundwork for detailed nuclear GFMC studies of
chiral EFT potentials for $A \leq 12$ nuclei, which will also impact
future simulations of larger systems with the auxiliary-field diffusion
Monte Carlo method.

\acknowledgments{We thank S.~Bacca, P.~Navr\'atil, and I.~Tews for
useful discussions.
The work of J.L., S.G., and J.C. is supported by the
U.S. Department of Energy, Office of Nuclear Physics, and by the NUCLEI
SciDAC program.
The work of S.G. is also supported by the LANL LDRD program.
The work of A.G. is supported by the Natural Sciences and Engineering
Research Council of Canada.
The work of A.S. is supported in part by ERC Grant No.~307986 STRONGINT,
the Helmholtz Alliance Program of the Helmholtz Association Contract 
No.~HA216/EMMI ``Extremes of Density and Temperature: Cosmic Matter in
the Laboratory,'' and by computing resources at the J\"ulich
Supercomputing Center.
The work of E.E. is supported in part by ERC Grant No.~259218 
NUCLEAREFT.
Computational resources have been provided by Los Alamos Open 
Supercomputing.
This research used also resources of the National Energy Research 
Scientific Computing Center (NERSC), which is supported by the Office of
Science of the U.S. Department of Energy under Contract 
No.~DE-AC02-05CH11231.

%
\end{document}